\documentclass[conference,a4paper]{IEEEtran}
\usepackage[left=1in, right=0.7in, bottom=1.4in, top=0.76in]{geometry}
\IEEEoverridecommandlockouts

\usepackage{cite}
\usepackage{amsmath,amssymb,amsfonts}
\usepackage{algorithmic}
\usepackage{graphicx}
\usepackage{textcomp}
\usepackage{xcolor}
\usepackage{fancyhdr}
\def\BibTeX{{\rm B\kern-.05em{\sc i\kern-.025em b}\kern-.08em
    T\kern-.1667em\lower.7ex\hbox{E}\kern-.125emX}}

\usepackage{subcaption}
\usepackage{caption}

\usepackage{subfig} 
\usepackage{subcaption} 

\usepackage{acronym}

\acrodef{ADCs}{analog-to-digital converters}
\acrodef{FEEC}{Faculty of Electrical Engineering and Communication}
\acrodef{BUT}{Brno University of Technology}
\acrodef{CIR}{channel impulse response}
\acrodef{RMS}{root mean square}
\acrodef{DS}{delay spread}
\acrodef{DAC}{digital-to-analog converters}
\acrodef{MMW}[MMW]{millimeter-wave}
\acrodef{FMCW}[FMCW]{frequency modulated continuous wave}
\acrodef{FSPL}[FSPL]{free space path loss}
\acrodef{FFT}[FFT]{fast fourier transform}
\acrodef{IFFT}[IFFT]{inverse fast fourier transform}
\acrodef{USB}[USB]{upper sideband}
\acrodef{LSB}[LSB]{lower sideband}
\acrodef{IRR}{image rejection ratio}
\acrodef{SNR}{signal to noise ratio}
\acrodef{MPC}{multipath component}
\acrodef{LOS}{line-of-sight}
\acrodef{NLOS}{non-line-of-sight}
\acrodef{GPS}{global positioning system}
\acrodef{TX}{transmitter}
\acrodef{RX}{receiver}
\acrodef{RF}{radio frequency}
\acrodef{ATT}{attenuator}
\acrodef{G}{gain}
\acrodef{RP}{relative power}
\acrodef{PDP}{power delay profile}
\acrodef{OWGA}{open waveguide antenna}
\acrodef{HPBW}{half power beam width}
\acrodef{AGV}{autonomous ground vehicles}
\acrodef{CTF}{channel transfer function}
\acrodef{V2V}{Vehicle-to-vehicle}
\acrodef{V2I}{vehicle-to-infrastructure}
\acrodef{SNR}{signal-to-noise ratio}
\acrodef{SSD}{solid-state drive}
\acrodef{I}{in-phase}
\acrodef{Q}{quadrature}
\acrodef{PDF}{probability density function}
\acrodef{CDF}{cumulative distribution function}
\acrodef{ITSs}{intelligent transportation systems}
\acrodef{ISAC}{integrated sensing and communication}
\acrodef{S-MPCs}{sensing multi-path components}
\acrodef{C-MPCs}{clutter multi-path components}
\acrodef{MCD}{\ac{MPC} distance}

\fancypagestyle{firstpage}
{
    \fancyhf{}
    \fancyhead[C]{The paper has been presented at the 2025 IEEE Conference on Antenna Measurements and Applications (CAMA), Antibes, France, November 18-20, 2025}    
    \fancyfoot[C]{This research was funded in part by the National Science Center (NCN), Poland, grant no. 2021/43/I/ST7/03294 (MubaMilWave). For this purpose of Open Access, the author has applied a CC-BY public copyright license to any Author Accepted Manuscript (AAM) version arising from this submission.}
}

\begin{document}
\makeatletter

\title{Vehicle-to-Vehicle Millimeter-Wave Channel Characterization at 60 and 80\,GHz}

\author{\IEEEauthorblockN{Radek Zavorka\IEEEauthorrefmark{1}, Jiri Blumenstein\IEEEauthorrefmark{1}, Tomas Mikulasek\IEEEauthorrefmark{1}, Josef Vychodil \IEEEauthorrefmark{1}, Hussein Hammoud\IEEEauthorrefmark{2},\\ Wojtuń Jarosław \IEEEauthorrefmark{4},Aniruddha Chandra \IEEEauthorrefmark{3}, Markus Hofer \IEEEauthorrefmark{6}, Jan M. Kelner \IEEEauthorrefmark{4}, Cezary Ziółkowski \IEEEauthorrefmark{4},\\ Christoph Mecklenbräuker \IEEEauthorrefmark{5}, Ales Prokes \IEEEauthorrefmark{1}}
\IEEEauthorblockA{\IEEEauthorrefmark{1}Department of Radio Electronics, Brno University of Technology, Brno, Czech Republic}
\IEEEauthorblockA{\IEEEauthorrefmark{2}University of Southern California, Los Angeles, USA}
\IEEEauthorblockA{\IEEEauthorrefmark{3}National Institute of Technology, Durgapur, India}
\IEEEauthorblockA{\IEEEauthorrefmark{4}
Institute of Communications Systems, Military University of Technology, Warsaw, Poland}
\IEEEauthorblockA{\IEEEauthorrefmark{5}Institute of Telecommunications, TU Wien, Vienna, Austria}
\IEEEauthorblockA{\IEEEauthorrefmark{6} AIT Austrian Institute of Technology, Vienna, Austria}
e-mail: xzavor03@vutbr.cz}

\maketitle
\thispagestyle{firstpage}

\begin{abstract}
This paper presents results from a vehicle-to-vehicle channel measurement campaign conducted in the millimeter-wave (MMW) frequency bands at center frequencies of 60\,GHz and 80\,GHz, each with a bandwidth of 2\,GHz. The measurements were performed in a dynamic oncoming-vehicle scenario using a time-domain channel sounder with high-resolution data acquisition. Power delay profiles were extracted to study the temporal evolution of multipath components, and the root mean square (RMS) delay spread was analyzed to characterize the temporal dispersion of the channel. The results demonstrate differences between the two frequency bands. At 60\,GHz, the RMS delay spread is well approximated by a Gaussian distribution with a higher median value, while at 80\,GHz it follows a lognormal distribution with a lower median. Furthermore, the number of resolvable multipath components was found to be nearly twice as high at 60\,GHz compared to 80\,GHz, highlighting the impact of antenna beamwidth and frequency-dependent propagation mechanisms. 
\end{abstract}

\begin{IEEEkeywords}
millimeter-wave, vehicle-to-vehicle, RMS delay spread, statistical distribution, channel sounder
\end{IEEEkeywords}

\section{Introduction}
\label{introduction}



\ac{V2V} communication in the \ac{MMW} bands is a key enabler of emerging \ac{ITSs} and autonomous driving, as it delivers the high data rates and low latency required for these applications \cite{V2Vsurvey}. This is achieved through the large bandwidth at \ac{MMW} frequencies, which supports gigabit-per-second links for cooperative awareness and real-time sensor sharing between vehicles.

In addition, the shorter wavelength at these frequencies allows for compact antenna arrays with high beamforming gain \cite{V2Vbeam}, thereby improving spatial selectivity and spectrum reuse. Despite these advantages, the propagation characteristics of \ac{V2V} channels at \ac{MMW} frequencies remain challenging. High path loss, susceptibility to blockage, and rapid temporal variations due to vehicle mobility all strongly impact link reliability.

Several measurement campaigns have been reported in the literature \cite{V2Vmmw1, V2Vmmw2}, focusing on different frequencies and scenarios within the \ac{MMW} bands. The authors in \cite{V2Xinvite} summarize results from multiple measurement campaigns carried out at 3.2\,GHz, 34.3\,GHz, and around 60\,GHz for both \ac{V2V} and \ac{V2I} scenarios, showing that in all analyzed cases the \ac{RMS} delay spread was below 100\,ns. In \cite{V2Vovertaking}, the authors report results from a \ac{V2V} campaign conducted in an urban street canyon at a center frequency of 60\,GHz with a bandwidth of 500~MHz. The \ac{PDP} was analyzed, and the study showed that a passenger car gives rise to a single multipath component, whereas a larger vehicle such as a truck produces several multipath components. 

The influence of blockage was investigated in \cite{V2Vblockage} at multiple frequency bands (6.75, 30, 60, and 73\,GHz) in both urban and highway scenarios. The results show that blockage loss strongly depends on the blocker size, increasing from 5.5–8.2\,dB for small blockers to 8.2–17\,dB for large blockers. Furthermore, \cite{V2VISAC} presents dynamic vehicular \ac{ISAC} channel measurements at 28\,GHz and introduces a tapped-delay line model that distinguishes between \ac{S-MPCs} and \ac{C-MPCs}. An \ac{MCD}-based tracking algorithm is employed to statistically characterize lifetimes, power evolution, clustering, and fading of \ac{MPC}, and the model is validated through comparison of measured and simulated results, confirming its suitability for vehicular \ac{ISAC} system design.


\subsection{Contribution of the Paper}
This paper presenting a dedicated \ac{V2V} channel measurement campaign conducted at 60\,GHz and 80\,GHz with 2\,GHz bandwidth. Using a time-domain channel sounder, we analyze \ac{PDP}, characterize the \ac{RMS} delay spread distributions, and study the temporal evolution of \ac{MPC} in an oncoming-vehicle scenario. The results provide empirical evidence of frequency-dependent differences between the two bands and highlight the importance of accurate, frequency-specific channel models for the design and evaluation of \ac{MMW} vehicular communication systems.

The main contributions of this paper are as follows:
\begin{itemize}
    \item Analysis of \ac{MPC} propagating between two oncoming vehicles at two \ac{MMW} frequency bands (60\,GHz and 80\,GHz), including their time-dependent statistical characterization.
    \item Evaluation and comparison of the \ac{RMS} delay spread characteristics and their statistical distributions at both frequencies.
    \item Investigation of the temporal evolution in the number of propagation paths and evaluation of differences between frequency bands.
\end{itemize}


\section{Measurement campaign description}

\begin{figure}[htbp]
    \centering
    \includegraphics[width=0.48\textwidth]{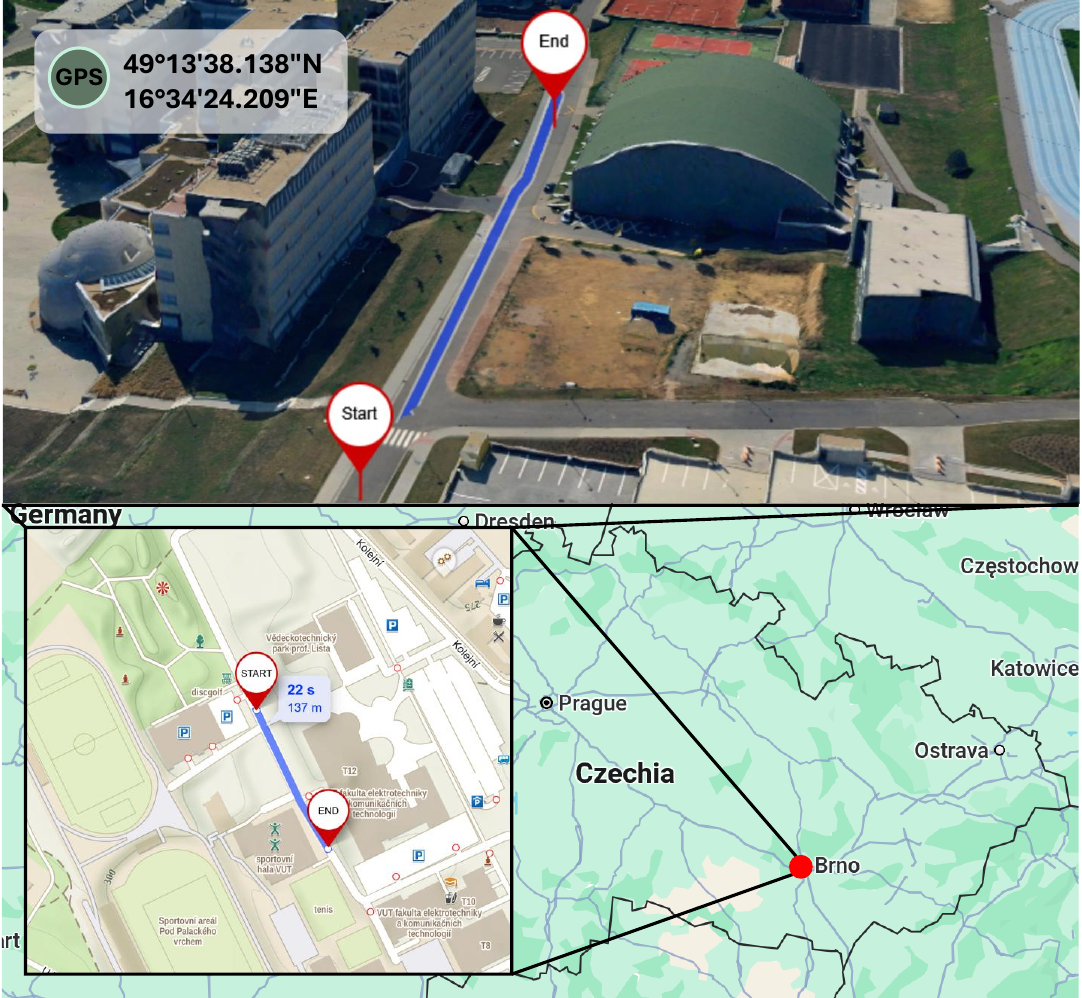}
    \caption{Location of the measurement campaign with a detailed 3D view.}
    \label{fig:maps}
\end{figure}

The measurement campaign focused on characterizing the high-mobility \ac{V2V} channel in a real-world environment. Measurement was conducted on the campus of the \ac{FEEC}, \ac{BUT}, Brno, Czech Republic, in the area between the T12 building on Technicka Street, the sports complex, and the adjacent parking lot. This location was selected because it combines open-space conditions with the presence of reflecting objects, creating a representative \ac{V2V} environment. A visualization of the campaign, including the route and surrounding environment, is shown in Fig.~\ref{fig:maps}. The studied scenario considered two vehicles approaching each other from opposite directions.

Along the road within the measurement area, several potential scatterers and reflectors were present, including parked cars, streetlights, traffic signs, medium-sized trees, and the nearby sports complex building. The pillars and wall of the building were located approximately 3\,m and 6.5\,m from the road, respectively, with the wall reaching a height of 9\,m. The road itself was approximately 6\,m wide, with sidewalks of about 2\,m on both sides.

\begin{figure}[htbp]
    \centering
    \includegraphics[width=0.48\textwidth]{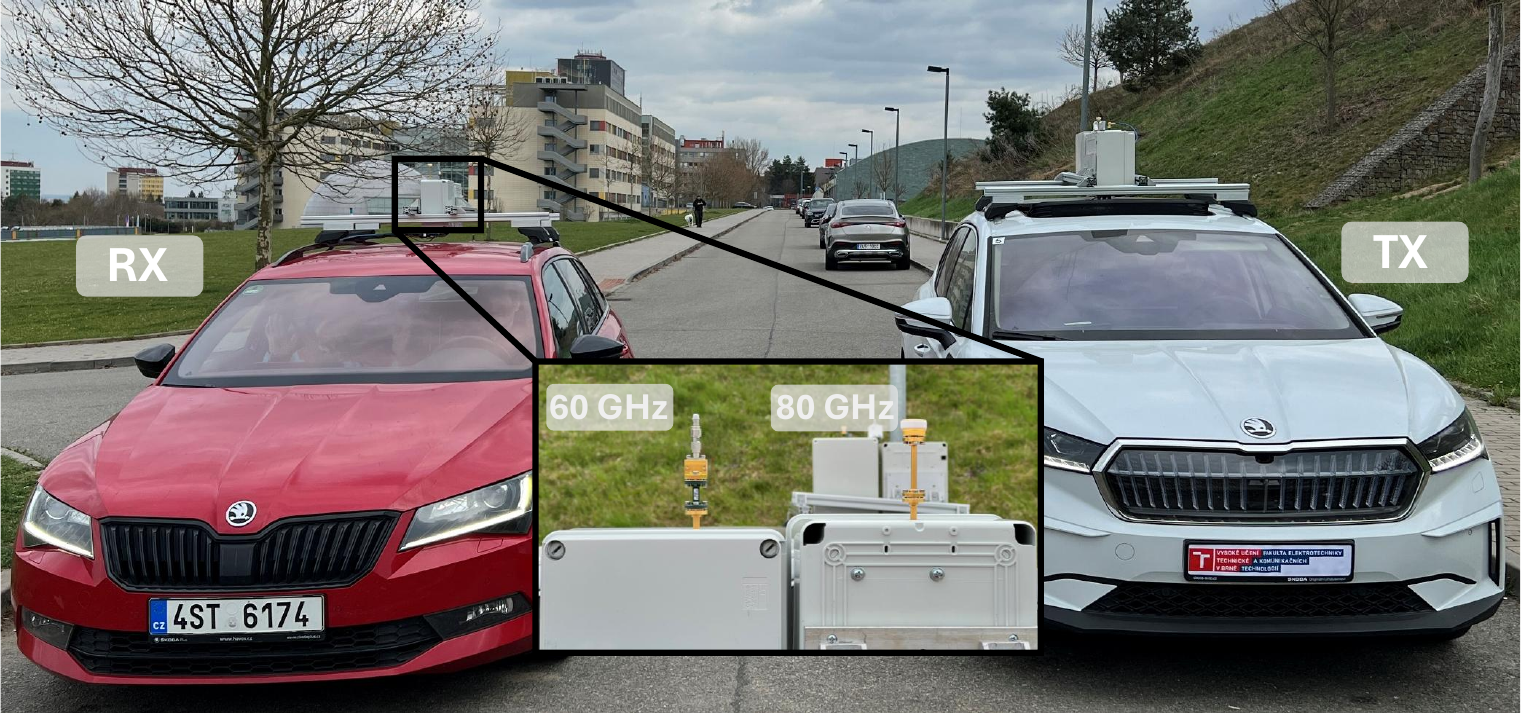}
    \caption{Vehicles equipped with antennas and front-end units on the roof for 60 and 80\,GHz.}
    \label{fig:cars}
\end{figure}

Transmit and receive antennas with front-end units were mounted on the roofs of the vehicles, as illustrated in Fig.~\ref{fig:cars}. The Skoda Eniaq carried both transmit antennas, with the 80\,GHz unit positioned closer to the front of the car and the 60\,GHz unit placed behind it toward the rear. The antennas were aligned at the same height, centered along the longitudinal axis of the vehicle, and separated by approximately 20~cm. The same configuration was applied at the receiver side, mounted on the second vehicle, a Skoda Superb III. The transmitter vehicle followed a trajectory from the defined start point to the end point at an average speed of about 45\,km/h, while the receiver vehicle traveled in the opposite direction at a speed of approximately 35\,km/h. The total \ac{CIR} recording spanned 8\,s, but for further processing only the most relevant 4\,s segment with the highest \ac{SNR} was selected.

\section{Measurement setup} 
The transmit baseband section of the time-domain channel sounder was realized using the Xilinx Zynq UltraScale+ RFSoC ZCU111 platform. High-speed \ac{DAC} with a sampling rate of 6.144\,GSPS generated the \ac{I} and \ac{Q} components of the intermediate-frequency signal. As a probing waveform, an \ac{FMCW} signal with both up- and down-chirp slopes was applied, ensuring a spectrum with minimal irregularities.

The main motivation for selecting an FMCW waveform was its robustness to system nonlinearities. With a bandwidth of $B = 2.048\,\mathrm{GHz}$ and a sweep duration of $T = 8\,\mu\mathrm{s}$, the waveform supports a high measurement rate of $f_{\mathrm{meas}} = 1/T = 125\,\mathrm{kHz}$ (measurements per second), while still preserving a satisfactory \ac{SNR}. In static measurement conditions, further improvement of the \ac{SNR} can be achieved through averaging.

\begin{figure}[htbp]
    \centering
    \includegraphics[width=\linewidth]{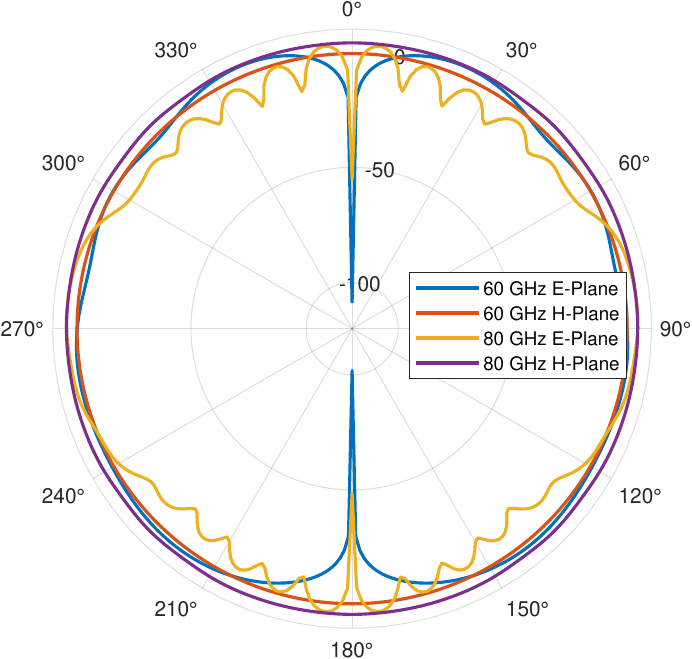}
    \caption{Comparison of antenna radiation patterns in the E-plane and H-plane at 60 and 80\,GHz.}
    \label{fig:ants}
\end{figure}

Upconversion to the target \ac{MMW} bands was carried out using Sivers IMA converters, specifically the FC1005V/00 for 60\,GHz and the FC1003E/03 for 80\,GHz. A frequency-stable, low phase-noise local oscillator signal was supplied by an Agilent 83752A generator. The resulting \ac{RF} signal was subsequently amplified by a QuinStar QPW-50662330-C1 amplifier at 60\,GHz or a Filtronic Cerus 4 AA015 amplifier at 80\,GHz, and then transmitted through an omnidirectional antenna. The radiation patterns of the 60\,GHz and 80\,GHz antennas are shown in Fig.~\ref{fig:ants}, and their detailed parameters are listed in Tab.~\ref{tab:ant_parameters}.

\begingroup
\setlength{\tabcolsep}{20pt} 
\begin{table}[h]
    \centering
    \caption{Parameters of omnidirectional antennas at 60 and 80\,GHz.}
    \begin{tabular}{l||c|c}
        &60\,GHz&80\,GHz\\
     \hline
      \hline
     Gain [dBi]&-0.3& 3.7\\
     \hline
     HPBW E-plane [°]&156&40\\

    \end{tabular}
    
    \label{tab:ant_parameters}
\end{table}
\endgroup

On the receive side, the architecture mirrored that of the transmitter. The signal, captured by the same type of omnidirectional antenna, was first amplified using either a QuinStar QLW-50754530-I2 amplifier at 60\,GHz or a Low Noise Factory $LNF\text{-}LNR55\_96WA\_SV$ amplifier at 80\,GHz. Downconversion was then carried out by a Sivers IMA FC1003V/01 (60\,GHz) or FC1003E/02 (80\,GHz) converter, driven by a local oscillator signal generated by an Agilent E8257D. The resulting intermediate-frequency signal, represented by its \ac{I} and \ac{Q} components, was digitized using high-speed \ac{ADCs} operating at 4.096\,GSPS on a second Xilinx Zynq UltraScale+ RFSoC ZCU111 board. The data were stored on an \ac{SSD} for subsequent offline processing. Additional information about the testbed, calibration procedure, and data processing for extracting the \ac{CIR} can be found in \cite{access_radek}.

\subsection{High-Resolution Data Recording Principle}
\begin{figure}[htbp]
    \centering
    \includegraphics[width=0.48\textwidth]{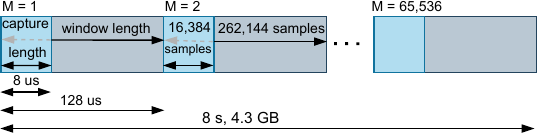}
    \caption{Graphical representation of the high-resolution data recording principle.}
    \label{fig:recording_principle}
\end{figure}

The measurement setup, based on the Xilinx Zynq UltraScale+ RFSoC ZCU111 board described above, enables high-resolution data acquisition. In each clock cycle, 256~bits (32~bytes) of data are captured, with every complex sample consisting of 4~bytes (16~bits for \ac{I} and 16~bits for \ac{Q}). This corresponds to 8 samples per cycle. Due to the large data volume, continuous recording over several seconds is not feasible for direct processing. Instead, the recording strategy is illustrated in Fig.~\ref{fig:recording_principle}. The acquisition window length was set to 262,144~samples, corresponding to $128\,\mu\mathrm{s}$. From each window, only the first 16,384~samples (equivalent to $8\,\mu\mathrm{s}$) were retained, while the remaining portion was discarded. To obtain a total duration of $8.4\,\mathrm{s}$, $M = 65,536$ windows were recorded, which required approximately 4.3\,GB of storage.

\section{Stochastic Channel Characterization}
This section presents stochastic channel characterization based on empirical \ac{V2V} channel measurements.  

Our time-domain channel sounder was used to estimate the \ac{CIR}, expressed as
\begin{equation}
\label{eq:CIR}
h(t, \tau) = \sum_{n=1}^{N(t)} \alpha_n(t)\,\delta\bigl(\tau - \tau_n(t)\bigr),
\end{equation}
where $h(t,\tau)$ is the equivalent complex baseband \ac{CIR}, $N(t)$ is the number of propagation paths, $\alpha_n(t)$ is the complex-valued path gain, and $\delta$ represents the Dirac delta function. The power distribution across delayed propagation paths is described by the \ac{PDP}, defined as

\begin{equation}
\label{eq:PDP}
P(\tau)= \mathbb{E}\{|h(t, \tau)|^2\},
\end{equation}

where $\mathbb{E}\{\cdot\}$ denotes the expectation taken over several realizations of \ac{CIR}s.

\begin{figure}[h]
    \centering
    \begin{subfigure}[b]{\linewidth}
        \centering
        \includegraphics[width=\linewidth]{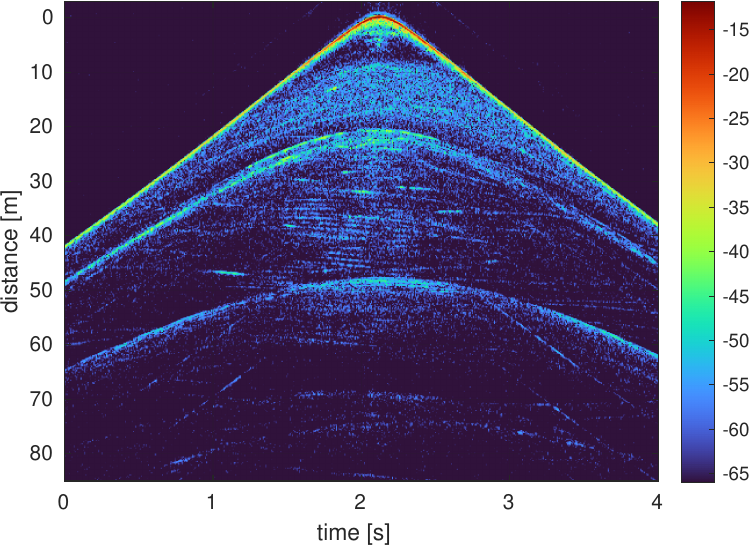}
        \caption{}
        \label{fig:pdp60}
    \end{subfigure}
    \vspace{1mm}
    \begin{subfigure}[b]{\linewidth}
        \centering
        \includegraphics[width=\linewidth]{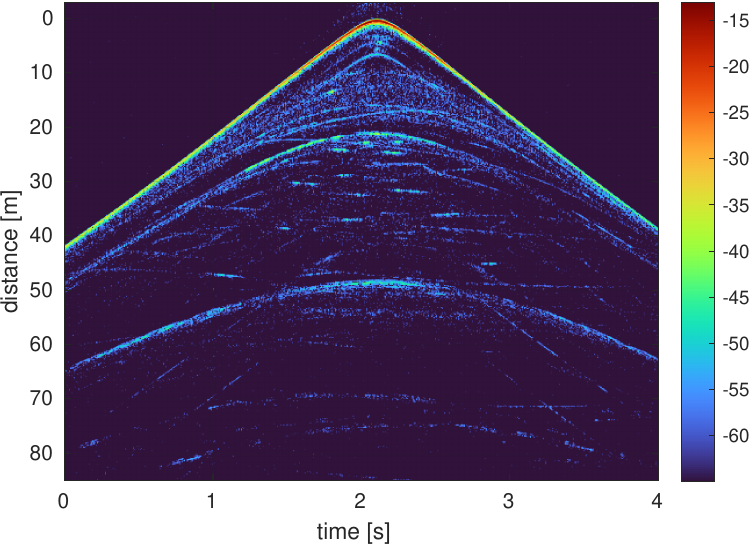}
        \caption{}
        \label{fig:pdp80}
    \end{subfigure}

    \caption{Measured \ac{PDP}s of passing vehicles for the 60\,GHz (a) and 80\,GHz (b) bands.}
    \label{fig:pdp60_80}
\end{figure}

The measured \ac{PDP} of the passing vehicles is shown in Fig.~\ref{fig:pdp60} and Fig.~\ref{fig:pdp80}. The dominant \ac{LOS} component is clearly visible, while the remaining \acp{MPC} originating from surrounding objects appear either in clusters or as isolated paths. In both cases, weak \acp{MPC} can be observed up to a distance of 80~m, which corresponds to a delay of 266.6\,ns. To improve the visibility of relevant components, all values below the noise floor were discarded. A fixed threshold of --66\,dB for the 60\,GHz band and --65\,dB for the 80\,GHz band was applied to suppress background noise. The use of different thresholds is attributed to slight differences in the measurement setup for each band, particularly in terms of dynamic range (40\,dB at 60\,GHz and 45\,dB at 80\,GHz). It should be noted that, despite careful calibration, small inconsistencies between the setups may still lead to variations in the observed \acp{MPC} in the \ac{PDP}. In particular, the distinct antenna \ac{HPBW} in both bands (156° at 60\,GHz and 40° at 80\,GHz) contributes to these differences.

\subsection{Frequency and Time Dependence of RMS Delay Spread}

The temporal dispersion of the radio channel is described by the \ac{RMS} delay spread. This parameter quantifies the extent of multipath propagation, where distinct signal components travel over different paths and arrive at the receiver with varying delays. The \ac{RMS} delay spread is defined as the square root of the second central moment of the \ac{PDP}:

\begin{equation}
\label{eq:RMS_delay_spread}
\sigma_{\tau}(t) = \sqrt{\overline{\tau^2}(t) - \left(\overline{\tau}(t)\right)^2}
\end{equation}

where $\overline{\tau}(t)$ denotes the mean excess delay and $\overline{\tau^2}(t)$ is the second moment of delay, both defined as

\begin{equation}
\label{eq:first_moment}
\overline{\tau}(t) = \frac{\sum_{n=1}^{N(t)} P(\tau_n,t)\,\tau_n(t)}{\sum_{n=1}^{N(t)} P(\tau_n,t)},
\end{equation}

\begin{equation}
\label{eq:second_moment}
\overline{\tau^2}(t) = \frac{\sum_{n=1}^{N(t)} P(\tau_n,t)\,\tau_n^2(t)}{\sum_{n=1}^{N(t)} P(\tau_n,t)}.
\end{equation}

Here, $P(\tau_n, t) = \mathbb{E}\{|h(t,\tau_n)|^2\}$ represents the \ac{PDP} tap at the delay~$\tau_n$, $N(t)$ is the number of multipath components at time $t$, and $h(t,\tau_n)$ denotes the complex \ac{CIR} at the delay~$\tau_n$ in time~$t$.

\begin{figure}[h]
    \centering
    \begin{subfigure}[b]{\linewidth}
        \centering
        \includegraphics[width=\linewidth]{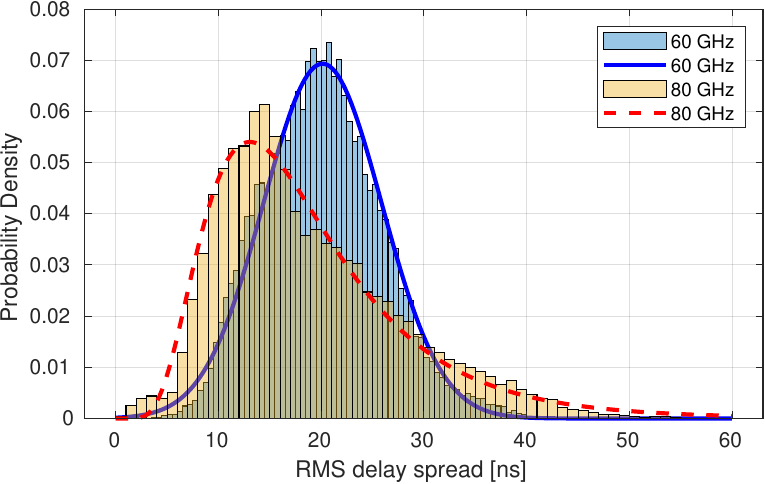}
        \caption{}
        \label{fig:pdf}
    \end{subfigure}
    \vspace{1mm}
    \begin{subfigure}[b]{\linewidth}
        \centering
        \includegraphics[width=\linewidth]{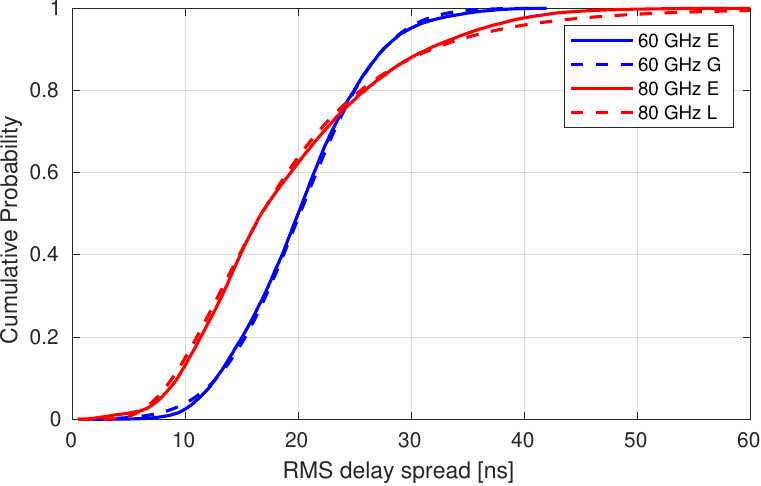}
        \caption{}
        \label{fig:cdf}
    \end{subfigure}

    \caption{Probability density function (a) and cumulative distribution function (b) of the RMS delay spread. Legend: Empirical (E), Gaussian (G), and Lognormal (L).}
    \label{fig:pdf+cdf}
\end{figure}

The statistical distribution of the \ac{RMS} delay spread over the measurement interval is shown in Fig.~\ref{fig:pdf}, where the \ac{PDF} for both frequency bands are depicted, while the corresponding \ac{CDF} are presented in Fig.~\ref{fig:cdf}. The \ac{PDF} illustrates the relative likelihood of particular \ac{RMS} delay spread values, whereas the \ac{CDF} represents the probability that the spread does not exceed a given threshold.
The results indicate that the two measured frequency bands exhibit distinct statistical behaviors. The distribution of the \ac{RMS} delay spread at 60\,GHz is well approximated by a Gaussian distribution, with the 50th percentile (median) equal to $19.97\,\mathrm{ns}$. In contrast, the 80\,GHz band is more accurately modeled by a lognormal distribution, with a corresponding median value of $16.71\,\mathrm{ns}$.

\begin{equation}
F_{\sigma_\tau}^{60}(\tau_{50}^{60}) = 0.5, \quad \tau_{50}^{60} = 19.97\,\mathrm{ns},
\end{equation}

\begin{equation}
F_{\sigma_\tau}^{80}(\tau_{50}^{80}) = 0.5, \quad \tau_{50}^{80} = 16.71\,\mathrm{ns},
\end{equation}

where $F_{\sigma_\tau}^{60}(\cdot)$ and $F_{\sigma_\tau}^{80}(\cdot)$ denote the \ac{CDF} of the \ac{RMS} delay spread at 60\,GHz and 80\,GHz, respectively.

\subsection{Temporal Variation of Multipath Components}

The complex propagation environment gives rise to a large number of \acp{MPC} between the transmitter and receiver. In realistic scenarios, the channel is inherently time-varying due to the motion of obstacles such as humans or vehicles. In our measurement campaign, the situation is even more dynamic since both the transmitter and the receiver were moving. As illustrated by the evolution of the \ac{PDP} in Fig.~\ref{fig:pdp60} and Fig.~\ref{fig:pdp80}, the vehicles started approximately 42\,m apart, approached each other, passed at around $t \approx 2.1$\,s, and then continued in opposite directions. Throughout the measurement, a strong \ac{LOS} component was present. The number of resolvable \acp{MPC} varied over time~$t$ within each \ac{PDP}, and additional fading effects were observed as a function of the propagation delay~$\tau_n$. These variations were determined by the relative positions of the vehicles, surrounding objects, and multipath reflections from the environment.

From the analysis of the \ac{PDP}, the number of resolvable propagation paths over time~$t$ was extracted for both frequency bands. Fig.~7(a) and Fig.~7(b) show the temporal variation and enable a comparison between the two bands. The overall shape of the curves reflects the scenario of two vehicles passing in opposite directions. At the beginning and end of the measurement, when the vehicles were relatively far apart, the number of resolvable paths was limited to about 10–20. As the vehicles approached each other, the number of paths increased almost linearly, reaching a sharp peak around $t \approx 2.1$\,s, and then decreased symmetrically as the vehicles moved away. The triangular shape of the curves highlights the strong dependence of multipath richness on the relative vehicle distance.


The 80\,GHz band is characterized by more pronounced short-term fluctuations around the moving average (red line), with a variance of approximately 12~paths, compared to about 5~paths at 60\,GHz. This behavior can be attributed to the shorter wavelength at 80\,GHz, which makes the channel more sensitive to small-scale variations in the environment. In contrast, the maximum number of resolvable paths was nearly twice as high at 60\,GHz, reaching about 80, whereas at 80\,GHz it peaked at only about 40. These differences arise not only from frequency-dependent propagation effects but also from distinct antenna radiation characteristics. In particular, the wider \ac{HPBW} of the antennas used at 60\,GHz enables the reception of a larger number of \ac{MPC}.

\begin{figure}[h]
    \centering
    \begin{subfigure}[b]{\linewidth}
        \centering
        \includegraphics[width=\linewidth]{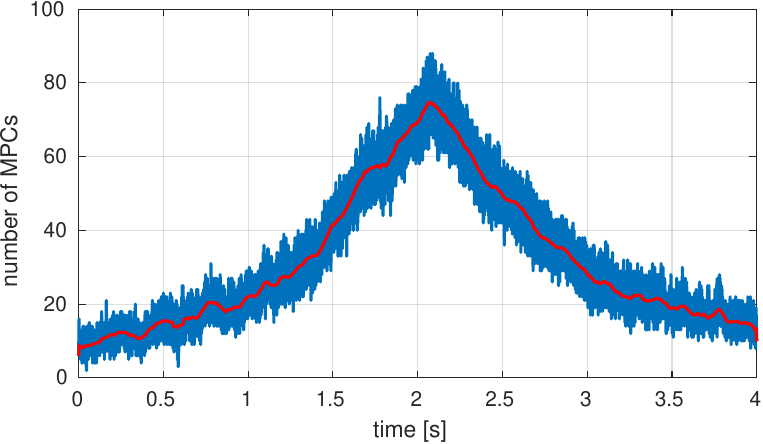}
        \caption{}
        \label{fig:nmpc60}
    \end{subfigure}
    \vspace{1mm}
    \begin{subfigure}[b]{\linewidth}
        \centering
        \includegraphics[width=\linewidth]{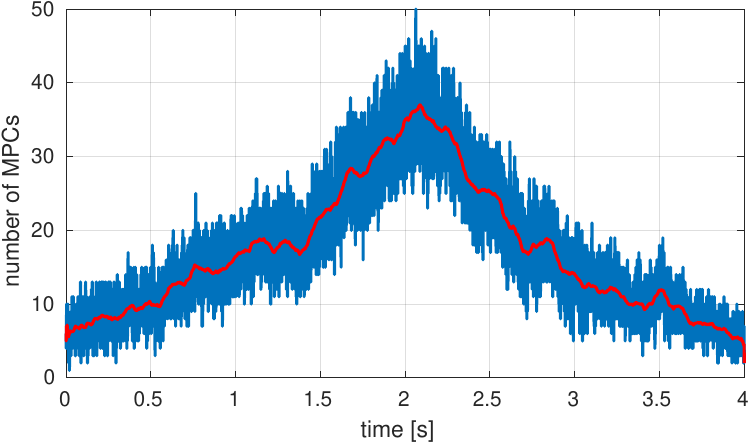}
        \caption{}
        \label{fig:nmpc80}
    \end{subfigure}

    \caption{Number of resolvable propagation paths over time – 60\,GHz (a) and 80\,GHz (b) bands.}
    \label{fig:nmpc}
\end{figure}

\section{Conclusion}
The results confirm that the 60\,GHz and 80\,GHz frequency bands exhibit distinct statistical distributions of the \ac{RMS} delay spread in the oncoming vehicle scenario, highlighting the frequency-dependent nature of multipath propagation in dynamic vehicular environments. At 60\,GHz, the delay spread is well approximated by a Gaussian distribution with a higher median value ($19.97\,\mathrm{ns}$), whereas at 80\,GHz the distribution follows a lognormal form with a lower median ($16.71\,\mathrm{ns}$). These findings indicate that within the \ac{MMW} spectrum, propagation characteristics cannot be generalized across frequency bands. The analysis of the temporal variation of multipath components further shows that the richness and stability of the channel are strongly dependent on both frequency and antenna characteristics. While the 60\,GHz band supports a larger number of resolvable propagation paths, partly due to the wider antenna \ac{HPBW}, the 80\,GHz band is more sensitive to small-scale changes in the environment, resulting in stronger fluctuations around the mean. 
Future work will focus on extending the analysis to additional vehicular scenarios to provide a more comprehensive understanding of high-frequency vehicular propagation channels.


\section*{Acknowledgment}
The research described in this paper was financed by the Czech Science Foundation, Project No. 23-04304L, Multi-band prediction of millimeter-wave propagation effects for dynamic and fixed scenarios in rugged time varying environments and by the Internal Grant Agency of the Brno University of Technology under project no. FEKT-S-23-8191. The work of A. Chandra is supported by the Chips-to-Startup (C2S) program no. EE-9/2/2021-R\&D-E from MeitY, GoI. The work of J. Kelner, J. Wojtuń and C. Ziółkowski was funded by the National Science Centre, Poland, under the OPUS-22 (LAP) call in the Weave program, as part of research project no. 2021/43/I/ST7/03294, acronym ‘MubaMilWave’.

\bibliographystyle{IEEEtran}
{\footnotesize
\bibliography{IEEEabrv,biblio_yousef.bib}
}


\end{document}